\documentclass[12pt]{article}
\usepackage[utf8]{inputenc}
\usepackage[british]{babel}
\usepackage{cmap}
\usepackage{lmodern}
\usepackage[T1]{fontenc}

\usepackage{amssymb, amsmath, amsthm}
\usepackage[a4paper,top=25mm,bottom=25mm,left=25mm,right=25mm]{geometry}
\usepackage{ragged2e}

\usepackage{authblk} 
\usepackage{pifont}
\usepackage{graphicx}
\usepackage[dvipsnames,svgnames,table]{xcolor}
\usepackage[figuresright]{rotating}
\usepackage{xtab} 
\usepackage{longtable} 
\usepackage{multirow}
\usepackage{footnote}
\usepackage[stable]{footmisc}
\usepackage{chngpage} 
\usepackage{pdflscape} 
\usepackage{tocbibind} 

\usepackage{pgfplots}
\pgfplotsset{compat=1.18}
\pgfplotsset{every tick label/.append style={font=\footnotesize}}
\usepackage{setspace}

\makesavenoteenv{tabular}
\usepackage{tabularx}
\usepackage{booktabs}
\usepackage{threeparttable} 
\usepackage[referable]{threeparttablex} 
\newcolumntype{R}{>{\raggedleft\arraybackslash}X}
\newcolumntype{L}{>{\raggedright\arraybackslash}X}
\newcolumntype{C}{>{\centering\arraybackslash}X}
\newcolumntype{A}{>{\columncolor{gray!25}}C}
\newcolumntype{a}{>{\columncolor{gray!25}}c}

\newlength{\tablen}

\usepackage{dcolumn} 
\newcolumntype{.}{D{.}{.}{-1}}

\usepackage{tikz}
\usetikzlibrary{arrows, calc, matrix, patterns, positioning, trees}
\usepackage[semicolon]{natbib}
\usepackage[hyphens]{url}
\usepackage{hyperref} 
\hypersetup{
  colorlinks   = true,    		
  urlcolor     = blue,    		
  linkcolor    = blue,			
  citecolor    = ForestGreen,		  	
}
\usepackage{microtype}
\usepackage[justification=centering]{caption} 

\usepackage[labelformat=simple]{subcaption}

\DeclareCaptionLabelFormat{parenthesis}{(#2)}
\makeatletter
\renewcommand\p@subfigure{\arabic{figure}.}
\makeatother

\DeclareCaptionLabelFormat{parenthesis}{(#2)}
\makeatletter
\renewcommand\p@subtable{\arabic{table}.}
\makeatother

\usepackage{alphalph}

\def\addlegendimage{\csname pgfplots@addlegendimage\endcsname}

\usepackage{enumitem}

\setlist[itemize]{leftmargin=2.5\parindent}
\setlist[enumerate]{leftmargin=2.5\parindent}

\theoremstyle{plain}

\theoremstyle{definition}

\theoremstyle{remark}

\def\keywords{\vspace{.5em} 
{\noindent \textit{Keywords}: }}

\def\AMS{\vspace{.5em} 
{\noindent \textbf{\emph{MSC} class}: }}

\def\JEL{\vspace{.5em} 
{\noindent \textbf{\emph{JEL} classification number}: }}

\title{Are penalty shootouts better than a coin toss? Evidence from international club football in Europe}
\author{
\href{https://sites.google.com/view/laszlocsato}{L\'aszl\'o Csat\'o}\thanks{~Corresponding author \newline
E-mail: \emph{laszlo.csato@sztaki.hun-ren.hu} \newline
Institute for Computer Science and Control (SZTAKI), Hungarian Research Network (HUN-REN), Laboratory on Engineering and Management Intelligence, Research Group of Operations Research and Decision Systems, Budapest, Hungary \newline 
Corvinus University of Budapest (BCE), Institute of Operations and Decision Sciences, Department of Operations Research and Actuarial Sciences, Budapest, Hungary}
$\qquad \qquad$
\href{https://sites.google.com/view/doragretapetroczy}{D\'ora Gr\'eta Petr\'oczy}\thanks{~E-mail: \emph{apetroczy@metropolitan.hu} \newline
Budapest Metropolitan University, Budapest, Hungary}
}
\date{\today}

\begin{document}

\newgeometry{top=25mm,bottom=25mm,left=25mm,right=25mm}
\maketitle
\thispagestyle{empty}

\begin{abstract}
\noindent
Penalty shootouts play a crucial role in the knockout stage of major football tournaments. Their importance has substantially increased from the 2021/22 season, when the Union of European Football Associations (UEFA) scrapped the away goals rule. Our paper examines whether the outcome of a penalty shootout can be predicted in UEFA club competitions. Based on all shootouts between 2000 and 2025, no evidence is found for the effect of the kicking order, the field of the match, or psychological momentum. In contrast to previous results, we do not detect any relationship between shootout success and relative team strength, quantified by differences in Elo ratings and the implied winning probability.
Thus, the hypothesis that penalty shootouts are close to a coin toss in international competitions for European football clubs cannot be rejected.
\end{abstract}


\AMS{62P20, 90B90}

\JEL{D79, Z20}

\clearpage
\restoregeometry

\section{Introduction} \label{Sec1}

The penalty shootout is the ultimate tie-breaking rule in (association) football and several other sports to decide which team qualifies in a knockout match. Penalty shootouts usually take place if the result is tied in both regular and extra time. Penalty shootouts can directly follow regular time, too: in the 2025 Leagues Cup, teams were awarded three points for a win in regular time, two points for a win by a penalty shootout, and one point for a loss by a penalty shootout \citep{LeaguesCup2025}.

As we demonstrate in Section~\ref{Sec2}, penalty shootouts are the subject of serious academic interest in the last one and a half decades. The present paper contributes to this discussion by investigating the following research question:
Does the shooting order, the field of the match, psychological momentum, or the strength of the teams affect the outcome of a penalty shootout?
All of these issues have partially been investigated in the existing literature as presented in Section~\ref{Sec2}.
However, our research has three innovative aspects.
First, we analyse a relatively homogeneous sample provided by matches played in UEFA club competitions without any matches from national cups, where the strengths of the opponents can vary widely.
Second, the psychological momentum enjoyed by the comeback team is identified in two ways, and one of them---which team scored the last goal?---is novel, even though it is also included in a paper written simultaneously with our study \citep{vanHemertvanderKamp2026}.
Third, team strength is measured by Football Club Elo Ratings, a well-established statistical method to create ratings based on past performance \citep{Aldous2017, GomesdePinhoZancoSzczecinskiKuhnSeara2024, vanEetveldeLey2019}. Previous studies used the division of the team \citep{ArrondelDuhautoisLaslier2019, Krumer2020b}, betting odds \citep{WunderlichBergeMemmertRein2020, Pipke2025}, or (relative) market values \citep{vanHemertvanderKamp2026} for this purpose.

In our sample the outcomes of the 268 penalty shootouts played in UEFA club competitions between 2000 and 2025, none of the variables above has a significant effect on winning. This contradicts the results of some previous works, which uncover a first-mover advantage \citep{ApesteguiaPalacios-Huerta2010, Palacios-Huerta2014, DaSilvaMioranzaMatsushita2018, RudiOlivaresShatty2020} and the positive impact of psychological momentum \citep{Krumer2021b}. Furthermore, according to our knowledge, \emph{all} existing studies agree that stronger teams are more likely to win in a shootout \citep{ArrondelDuhautoisLaslier2019, Krumer2020b, WunderlichBergeMemmertRein2020, Pipke2025, vanHemertvanderKamp2026}, which cannot be observed in UEFA club competitions using Football Club Elo Ratings. A possible reason is the smaller strength difference in UEFA club competitions compared to national cups, which may limit the detectable effect of strength. Naturally, our finding cannot be interpreted as ``ability does not matter'': the analysis is conditional on matches reaching a penalty shootout, and stronger teams have recently been demonstrated to have a higher probability of qualifying in UEFA club competitions \citep{Csato2024c}.

Our findings could be especially important for the Union of European Football Associations (UEFA) when they decide on rule changes in the future.
In 2021, UEFA abolished the so-called away goals rule, which awarded the prize (qualification for the next round) in a two-legged match to the team that scored more away goals. This change has increased the probability of reaching a penalty shootout by approximately ten percentage points, from below 5\% to nearly 15\%, in an \emph{ex ante} perfectly balanced tie \citep{ForrestKosciolekTena2025}.
In addition, UEFA has recently given serious consideration to removing extra time in order to ease the burden of the best players \citep{Ames2025}. Since the outcome of a penalty shootout appears to be close to a coin toss in UEFA club competitions with respect to these observables, such a measure might favour weaker teams and teams playing away that, contrary to extra time (see \citet{Bahamonde-BirkeBahamonde-Birke2023} for the latter effect), are not disadvantaged in the shootout, level the playing field, and decrease the dominance of the top clubs.

The paper is structured as follows.
An overview of related studies is given in Section~\ref{Sec2}. The data and the methodology are described in Sections~\ref{Sec3} and \ref{Sec4}, respectively. Section~\ref{Sec5} presents our results, while Section~\ref{Sec6} offers their discussion and some concluding remarks.

\section{Related literature} \label{Sec2}

The issue of first-mover advantage in football penalty shootouts has received serious attention since the pioneering work of \citet{ApesteguiaPalacios-Huerta2010}. Some studies find evidence for a significant advantage enjoyed by the team kicking the first penalty in each round \citep{ApesteguiaPalacios-Huerta2010, Palacios-Huerta2014, DaSilvaMioranzaMatsushita2018, RudiOlivaresShatty2020}. These results have inspired formal modelling of the first-mover advantage, as well as several proposals for alternative mechanisms to mitigate or even eliminate this source of unfairness \citep{Palacios-Huerta2012, Echenique2017, BramsIsmail2018, VandebroekMcCannVroom2018, DelGiudice2019, AnbarciSunUnver2021, Csato2021c, Csato2021a, CsatoPetroczy2022a, BramsIsmailKilgour2024, HuangLiangDaiPollock2026}. The alternating (or ABBA) order was even tried in various tournaments between 2017 and 2018 \citep{Palacios-Huerta2020, Csato2021c}, albeit the IFAB (International Football Association Board), the rule-making body of football, stopped the experiments in 2018 \citep{FIFA2018b}. It is worth noting that the first-mover advantage seems to be non-existent in tennis tiebreaks, which use the ABBA sequence \citep{Cohen-ZadaKrumerShapir2018, DaSilvaMioranzaMatsushita2018}.  

On the other hand, several papers report no significant difference between the winning probability of the first-mover and the second-mover \citep{KocherLenzSutter2012, ArrondelDuhautoisLaslier2019, Santos2023, vanHemertvanderKamp2026}. According to \citet{KassisSchmidtSchreyerSutter2021}, teams whose captains win the coin toss and can choose the shooting order enjoy the advantage.
These conflicting results can partially be attributed to limited sample sizes. According to the simulation exercise of \citet{VandebroekMcCannVroom2018}, a sample of 540 shootouts with a fixed underlying mathematical model produces considerable variation: the winning proportion of a team that enjoys a significant advantage in the model becomes significant in only 56.5\% of the 10 thousand simulations at 5\% level of significance.

Two recent studies \citep{VollmerSchochBrandes2024, Pipke2025} analysing the highest number of penalty shootouts with respect to first-mover advantage---1759 and 7116, respectively---do not find evidence for the effect of the shooting sequence. However, while the dataset of \citet{Pipke2025} includes all available shootouts up to the 2023/24 season, this does not mean that first-mover advantage cannot emerge in a more homogeneous subset of matches.

Regarding the existence of a penalty-specific home advantage, \citet{ApesteguiaPalacios-Huerta2010} show no difference in the chances of home and away teams for 129 shootouts between 1976 and 2003. \citet{KocherLenzSutter2012} reinforce this based on 540 shootouts between 1970 and 2003. Analogously, playing at home does not influence the probability of winning in 252 shootouts from French cup competitions \citep{ArrondelDuhautoisLaslier2019}.
The results of \citet{WunderlichBergeMemmertRein2020} suggest no home advantage in 1067 penalty shootouts that took place between 2004 and 2018. The logistic regressions of \citet{Bahamonde-BirkeBahamonde-Birke2023} and \citet{vanHemertvanderKamp2026}, based on 471 and 409 shootouts, respectively, demonstrate the absence of such an effect, too.
Even though \citet{vanOursvanTuijl2024} state that a penalty shootout does not seem to be a lottery in the Dutch domestic cup, the 20 home wins out of the 33 shootouts (61\%) are not significantly different from 50\%.

We know only two studies on the impact of psychological momentum, although the performance of men is significantly affected by psychological momentum in sports \citep{Cohen-ZadaKrumerShtudiner2017}.
\citet{Krumer2021b} analyses 214 penalty shootouts from two-legged matches in European cups (European Champion Clubs' Cup/UEFA Champions League, European Cup Winners' Cup/UEFA Cup Winners' Cup, UEFA Cup/UEFA Europa League) between 1970 and 2018. Every additional goal scored by the away team in regular time in the second leg increases its probability of winning the penalty shootout by 12.5 percentage points. However, if the home team wins in regular time, it has only a 50\% chance to qualify after a shootout.
\citet{vanHemertvanderKamp2026} consider 409 penalty shootouts between the 2008/09 to 2023/24 seasons, from national competitions (cups, supercups, playoffs for promotion, relegation, qualification) in eight European countries, as well as from international club competitions and tournaments for national teams. Both equalising (the randomly assigned Team A scored the last goal) and conceding (Team A conceded the last goal) dummies are insignificant in their logistic regressions.

\citet{ArrondelDuhautoisLaslier2019} reveal that the team playing at a higher level than its opponent wins the penalty shootout with an approximately 20\% higher probability. Analogously, a team from a higher division has a significantly higher chance to win according to \citet{Krumer2020b}: a difference in one league translates to a gap of 8 percentage points in winning probabilities. For instance, a team from the first division defeats its opponent from the second division with a probability of 54\%. The author studies 586 shootouts in the cup competitions of the top five European football nations (England, France, Germany, Italy, Spain) between 1979 and 2018.

Stronger football teams have a significantly higher chance of winning a penalty shootout based on 1067 shootouts from the domestic cup competitions of ten European associations (the top five, as well as Belgium, Portugal, Russia, Turkey, Ukraine), the Brazilian cup, the UEFA Champions League, and the UEFA Europa League \citep{WunderlichBergeMemmertRein2020}. However, the effect of team strength on success remains rather small; the winning probability does not exceed 60\% even against an extremely weak team. The novelty of \citet{WunderlichBergeMemmertRein2020} resides in the measure of strength difference, determined by winning probabilities based on pre-match betting odds, adjusted for home advantage. 
The recent analysis of all available penalty shootouts until the 2023/24 season reinforces that the favourites identified by pre-match odds are more likely to win \citep[Table~A.5]{Pipke2025}. But the favourite team might not be the stronger team due to home advantage, as remarked by \citet{WunderlichBergeMemmertRein2020}.

\citet{BrinkschulteWunderlichFurleyMemmert2023} examine the effect of skill on the success of individual penalty kicks. Skill is a composite measure of player market value, goalkeeper market value, and team quality derived from betting odds. Based on 1574 penalties in major international tournaments, the probability that the goalkeeper saves a penalty that is shot on target decreases if a highly skilled player takes the kick.

\citet{vanHemertvanderKamp2026} investigate the effect of team strength quantified by the market value of the starting 11 and the average market value of the players, including the goalkeeper,
who participated in the shootout. Based on 409 observations, the winning percentages gradually increase for stronger teams; teams in quartile 4 with the highest relative market value win with a probability of almost 60\%. This association is reinforced by logistic regressions: if the market value of a team relative to its opponent increases by 1\%, the odds of winning the shootout are higher by 1.3\%. On the other hand, none of the covariates---team age, playing at home/away, scoring the last goal, being the first-mover---has a significant impact on winning.

To summarise, \emph{all} previous studies considering the strength of teams have found at least some bias towards the stronger team in winning penalty shootouts. In other words, penalty shootouts are not equivalent to a pure coin toss; ability is an important success factor even for a mechanical task carried out in a high-pressure environment.

\section{Data} \label{Sec3}

\begin{table}[t!]
  \centering
  \caption{Penalty shootouts in UEFA club competitions from the 2000/01 season to 2025}
  \label{Table1}
\centerline{
\begin{threeparttable}
    \rowcolors{1}{}{gray!20}
    \begin{tabularx}{1.15\textwidth}{lCCCC} \toprule
          & Champions League & Europa League & Conference League & Total \\ \bottomrule
    Qualification & 41 (19) & 103 (31) & 63 (63) & 207 (113) \\
    Knockout stage & 14 (5) & 26 (8) & 10 (10) & 50 (23) \\
    Final & 6 (0) & 5 (3) & 0 (0) & 11 (3) \\ \toprule
    Total & 61 (24) & 134 (42) & 73 (73) & 268 (139) \\ \bottomrule
    \end{tabularx}
\begin{tablenotes} \footnotesize
\item
UEFA Conference League was called UEFA Europa Conference League until the 2023/24 season.
\item
Knockout stage stands for the knockout stage without the final. The knockout stage contains two-legged matches, except for the final, which is played on a neutral field.
\item
Numbers in parenthesis show the matches played from the 2020/21 season to 2025.
\end{tablenotes}
\end{threeparttable}
}
\end{table}

We have collected data from all penalty shootouts that took place in UEFA club competitions, including their qualifiers, from the 2000/01 season until the end of 2025. In the 2025/26 season, only the qualification matches are considered since the knockout stage of this season is played only in 2026.
Table~\ref{Table1} summarises the distribution of shootouts across the three tournaments, as well as across their stages. The final is distinguished from other knockout stage matches because it consists of one game played at a neutral venue and is not the usual two-legged match, where one team plays at home in a second leg.

\begin{figure}[t!]
\centering

\begin{tikzpicture}
\begin{axis}[
width = \textwidth, 
height = 0.6\textwidth,
xmajorgrids,
ymajorgrids,
xlabel = {Season},
xlabel style = {align=center, font=\small},
x tick label style={/pgf/number format/.cd,set thousands separator={}},
ylabel = {Number of penalty shootouts},
ylabel style = {align=center, font=\small},
xmin = 1999.5,
xmax = 2025.5,
ybar,
bar width = 6pt,
ymin = 0,
]
\addplot [blue, thick, pattern = crosshatch dots, pattern color = blue] coordinates{
(2000,2)
(2001,5)
(2002,6)
(2003,7)
(2004,6)
(2005,4)
(2006,3)
(2007,13)
(2008,7)
(2009,4)
(2010,6)
(2011,7)
(2012,10)
(2013,10)
(2014,8)
(2015,5)
(2016,10)
(2017,3)
(2018,6)
(2019,7)
(2020,21)
(2021,19)
(2022,31)
(2023,25)
(2024,29)
(2025,14)
};
\end{axis}
\end{tikzpicture}

\captionsetup{justification=centerfirst}
\caption{Number of penalty shootouts in UEFA club competitions in each season \\ \vspace{0.2cm}
\footnotesize{\emph{Note}: Seasons are identified by their starting year.}}
\label{Fig1}
\end{figure}
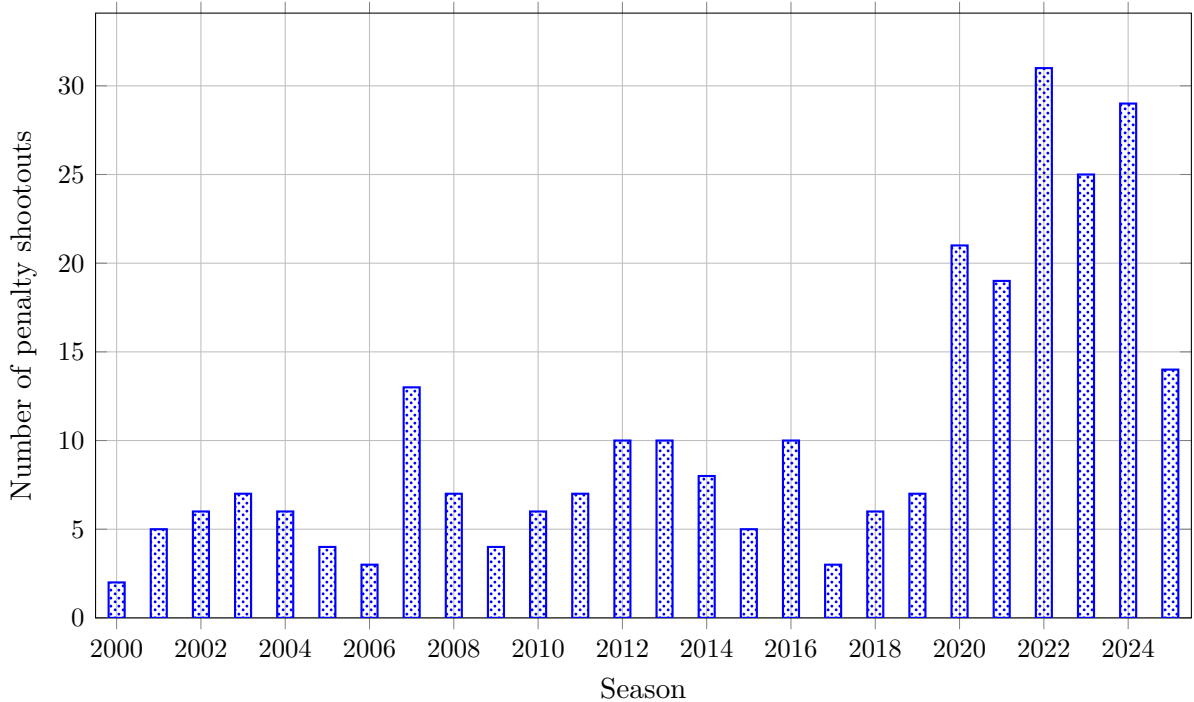


Figure~\ref{Fig1} presents the number of penalty shootouts in each season from 2000/01 to 2025/26 (in the latter case, without the spring of 2026). Besides the natural fluctuation, a dramatic increase can be seen from 2020/21, when most qualification matches were played in one leg instead of two due to COVID-19 restrictions, such that the field of the match was decided by a random draw. In addition, the away goals rule was abolished, and a new competition, the UEFA (Europa) Conference League, started in 2021/22. Therefore, our sample is dominated by recent matches over these 25 years: 139 of the 268 penalty shootouts (51.9\%) took place in the last six seasons (Table~\ref{Table1}).

For all matches decided by a penalty shootout, the dataset contains
(a) the time of the match;
(b) the round of the game in the tournament;
(c) the names of both teams;
(d) the location of the match;
(e) the number of goals scored by both teams in the match;
(f) the name of the team that scored the last goal in regular or extra time (if relevant);
(g) the team that kicked the first penalty (first-mover).

The first-mover advantage can be analysed on the basis of all (268) penalty shootouts since this information is always available.
On the other hand, some matches were played on a neutral field. In order to examine home advantage, we consider two sets of matches, an extended and a restricted set, where the latter is a strict subset of the former. The extended set does not contain the finals except for the 2011/12 UEFA Champions League final, which was played in M\"unchen by Bayern M\"unchen and Chelsea (10 matches are dropped). In addition, one match played by an Israeli and two matches played by Belarusian teams in 2024 and 2025 are disregarded, which leads to 255 matches in the extended set. In the restricted set, 20 matches played behind closed doors in the 2020/21 season, as well as the 2011/12 UEFA Champions League final, are removed.

We identify a comeback team in two different ways. The first adopts the method of \citet{Krumer2021b}: the team that scored more goals than its opponent (has a positive goal difference) in the match is the comeback team. Consequently, 121 matches where the second leg game is tied should be ignored.
The second approach is novel: the team that scored the last equalising goal, which is responsible for reaching the penalty shootout, is the comeback team. Hence, only 37 shootouts could not be examined.

The strengths of teams are quantified by Football Club Elo Ratings (\href{http://clubelo.com}{http://clubelo.com}). \citet{Csato2024c} shows that this measure of strength predicts the results of the UEFA Champions League more accurately than the official UEFA club coefficient. Unsurprisingly, Football Club Elo Ratings are widely used in the literature, too \citep{BoskerGurtler2024, Csato2022b, YildirimBilman2025a, YildirimBilman2025b}. They are available for all penalty shootouts, although we have found a problem in the database of Football Club Elo Ratings (see Appendix~\ref{Sec_A1} for details). Thus, the impact of the difference in team strength is studied based on 266 penalty shootouts in the baseline, because two Elo ratings are inaccurate.
As a robustness check, we also use different thresholds $t$: a match is retained in the sample only if the difference between the Elo ratings exceeds $t$.

It may be argued that betting odds could be a better predictor of team strength in penalty shootouts.
However, betting odds are known to suffer from several biases, such as systematic distortions in the pricing of own national teams \citep{BraunKvasnicka2013}, the favourite-longshot bias (favourites win more often and longshots win less often than the market probabilities imply) \citep{CainLawPeel2000}, higher prices for bets on teams with relatively more Facebook Likes \citep{FeddersenHumphreysSoebbing2017}, and slow reaction to disappeared home advantage \citep{WinkelmannDeutscherOtting2021}.

Furthermore, the betting odds published by bookmakers are usually related to the outcome of one particular match, but our database mostly contains the second legs of knockout ties. Therefore, the favourite team identified by betting odds might not be the stronger team because of home advantage. Even though \citet[Section~2.3]{WunderlichBergeMemmertRein2020} offers a method to adjust this bias, which allows the classification of a slight favourite as the slightly weaker team, such a procedure is not necessarily reliable in two-legged matches. The reason is the incentives of the teams: one team could be satisfied by a small loss if it still implies its qualification. For instance, if the away team won the first match by two goals, it does not have powerful incentives to attack in the second match even if it has scored one goal less than the other team. The betting odds of a home/away win certainly reflect the effect of these incentives, which have substantially changed from the 2021/22 season due to the abolition of the away goals rule.

Compared to Elo ratings, betting odds can indeed reflect better the current form of the teams by taking, for example, injuries, into account. However, it is far from clear that these factors are relevant to penalty shootouts, where success is based on individual skills, not on collaboration within a team.
The use of Elo ratings, given by an ``objective'' mathematical formula, may also be more persuasive for decision-makers than betting odds driven by market sentiment.
Finally, Football Club Elo Ratings have recently been demonstrated to outperform UEFA club coefficients, the official measure of team strength in these competitions, with respect to predictive accuracy \citep{Csato2024c}.
Based on these arguments, we follow the approach of previous papers \citep{ArrondelDuhautoisLaslier2019, Krumer2020b, WunderlichBergeMemmertRein2020, Pipke2025, vanHemertvanderKamp2026}, and do not consider alternative measures of team ability.

\section{Methodology} \label{Sec4}

All research questions are first investigated by a two-sided binomial test.
The number of penalty shootout wins and losses is calculated for the first-mover, the home team (two cases depending on the definition of neutral field), the comeback team (two cases depending on the type of psychological momentum), and the stronger team (five cases depending on the minimal threshold $t$).

The connection between the relative strength of teams and the probability of winning the penalty shootout is assessed by logistic regressions, too. The binary dependent variable is the home win/loss. Therefore, matches played on a neutral venue are removed due to the small sample size, similar to \citet{WunderlichBergeMemmertRein2020}.
Denote the Elo rating of the home and the away team by $E_h$ and $E_a$, respectively. The independent variable $\Delta_{ha}$ is either the difference of Elo ratings divided by 100 ($\Delta_{ha}^{(1)} = \left( E_h - E_a \right) /100$), or the winning probability of the home team implied by the Elo ratings according to the Elo equation of Football Club Elo Ratings (\href{http://clubelo.com/system}{http://clubelo.com/system}):
\[
\Delta_{ha}^{(2)} = \frac{1}{1 + 10^{- \left( E_h - E_a \right) / 400}}.
\]
Note that $\Delta_{ah}^{(1)} = - \Delta_{ha}^{(1)}$ and $\Delta_{ah}^{(2)} = 1 - \Delta_{ha}^{(2)}$.
The Elo difference is divided by 100 in order to remove the uninformative digits from the estimated coefficients.

\begin{figure}[t!]
\centering

\begin{subfigure}{\textwidth}
\centering
\caption{Distribution of strength difference $\Delta^{(1)}$}
\label{Fig2a}

\begin{tikzpicture}
\begin{axis}[
width = \textwidth, 
height = 0.5\textwidth,
xmajorgrids,
ymajorgrids,
xlabel = {Home team Elo rating minus away team Elo rating},
xlabel style = {align=center, font=\small},
symbolic x coords = {$-$700--$-$650,$-$650--$-$600,$-$600--$-$550,$-$550--$-$500,$-$500--$-$450,$-$450--$-$400,$-$400--$-$350,$-$350--$-$300,$-$300--$-$250,$-$250--$-$200,$-$200--$-$150,$-$150--$-$100,$-$100--$-$50,$-$50--0,0--50,50--100,100--150,150--200,200--250,250--300,300--350,350--400,400--450,450--500},
xtick=data,
enlarge x limits = 0.02,
x tick label style = {rotate=90},
ybar,
bar width = 8pt,
ymin = 0,
scaled y ticks = false,
yticklabel style = {/pgf/number format/fixed,/pgf/number format/precision=5},
ylabel = {Number of penalty shootouts},
ylabel style = {align=center, font=\small},
]
\addplot [blue, thick, pattern = crosshatch dots, pattern color = blue] coordinates{
($-$700--$-$650,1)
($-$650--$-$600,0)
($-$600--$-$550,0)
($-$550--$-$500,0)
($-$500--$-$450,0)
($-$450--$-$400,2)
($-$400--$-$350,5)
($-$350--$-$300,8)
($-$300--$-$250,9)
($-$250--$-$200,11)
($-$200--$-$150,23)
($-$150--$-$100,27)
($-$100--$-$50,26)
($-$50--0,26)
(0--50,27)
(50--100,33)
(100--150,22)
(150--200,16)
(200--250,10)
(250--300,12)
(300--350,4)
(350--400,3)
(400--450,0)
(450--500,1)
};

\coordinate (A) at (axis cs:{[normalized]13.5},0);
\coordinate (B) at (axis cs:{[normalized]13.5},\pgfkeysvalueof{/pgfplots/ymax});

\draw [black,very thick] (A) -- (B);
\end{axis}
\end{tikzpicture}
\end{subfigure}

\vspace{0.25cm}
\begin{subfigure}{\textwidth}
\centering
\caption{Distribution of strength difference $\Delta^{(2)}$}
\label{Fig2b}

\begin{tikzpicture}
\begin{axis}[
width = \textwidth, 
height = 0.5\textwidth,
xmajorgrids,
ymajorgrids,
xlabel = {Winning probability of the home team},
xlabel style = {align=center, font=\small},
symbolic x coords = {0--0.05,0.05--0.1,0.1--0.15,0.15--0.2,0.2--0.25,0.25--0.3,0.3--0.35,0.35--0.4,0.4--0.45,0.45--0.5,0.5--0.55,0.55--0.6,0.6--0.65,0.65--0.7,0.7--0.75,0.75--0.8,0.8--0.85,0.85--0.9,0.9--0.95,0.95--1},
xtick=data,
enlarge x limits = 0.02,
x tick label style = {rotate=90},
ybar,
bar width = 8pt,
ymin = 0,
scaled y ticks = false,
yticklabel style = {/pgf/number format/fixed,/pgf/number format/precision=5},
ylabel = {Number of penalty shootouts},
ylabel style = {align=center, font=\small},
]
\addplot [blue, thick, pattern = crosshatch dots, pattern color = blue] coordinates{
(0--0.05,1)
(0.05--0.1,3)
(0.1--0.15,11)
(0.15--0.2,11)
(0.2--0.25,14)
(0.25--0.3,21)
(0.3--0.35,24)
(0.35--0.4,18)
(0.4--0.45,19)
(0.45--0.5,16)
(0.5--0.55,19)
(0.55--0.6,25)
(0.6--0.65,17)
(0.65--0.7,20)
(0.7--0.75,16)
(0.75--0.8,9)
(0.8--0.85,14)
(0.85--0.9,7)
(0.9--0.95,1)
(0.95--1,0)
};

\coordinate (A) at (axis cs:{[normalized]9.5},0);
\coordinate (B) at (axis cs:{[normalized]9.5},\pgfkeysvalueof{/pgfplots/ymax});

\draw [black,very thick] (A) -- (B);
\end{axis}
\end{tikzpicture}
\end{subfigure}

\caption{Distribution of the difference between team strengths}
\label{Fig2}

\end{figure}
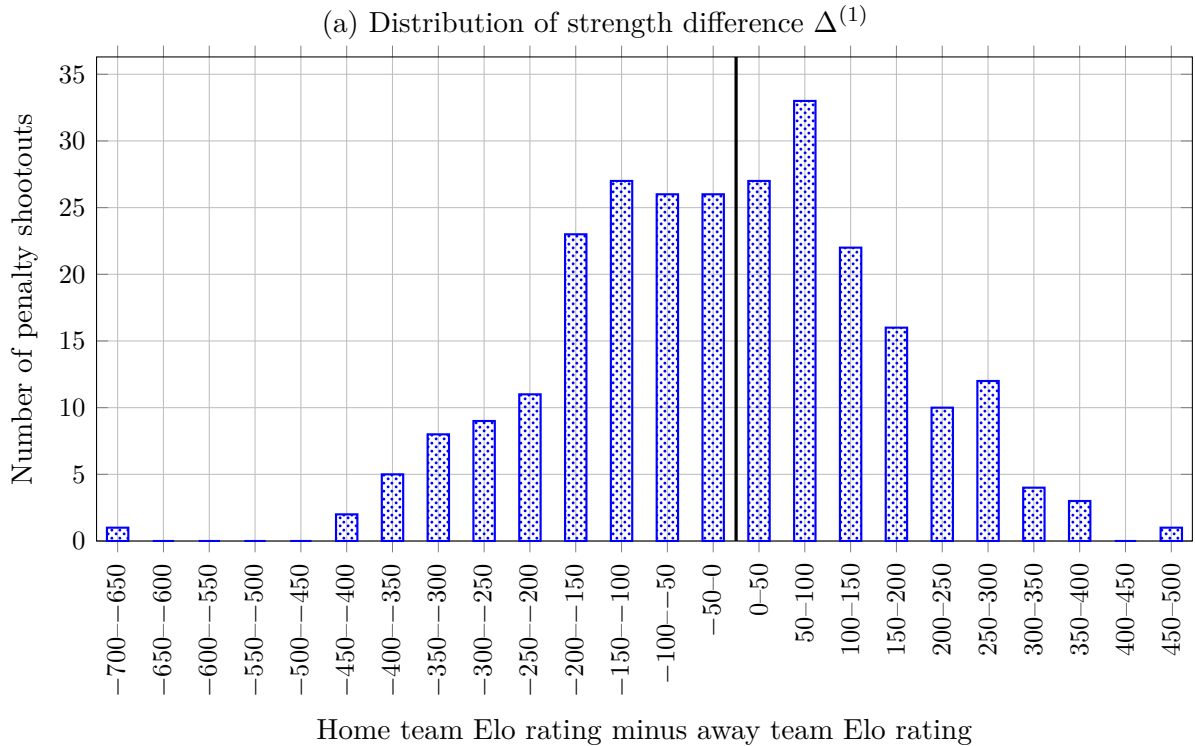
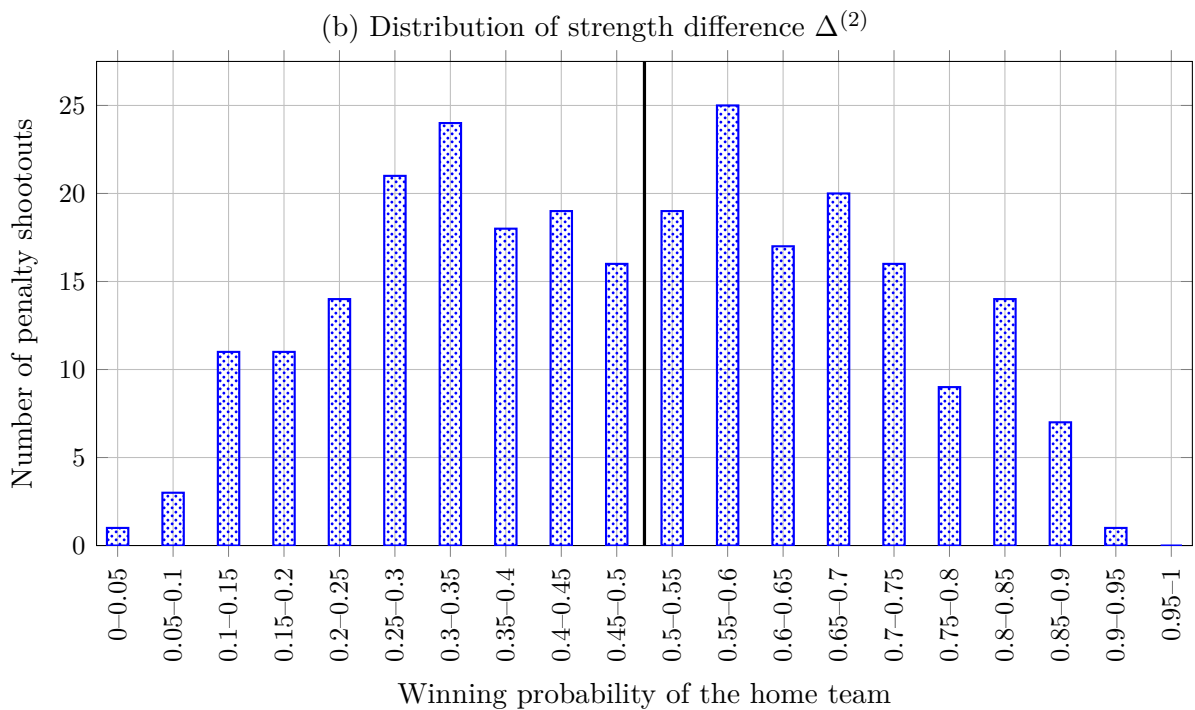


Figure~\ref{Fig2} shows histograms of strength difference $\Delta$: Figure~\ref{Fig2a} if the difference is measured directly by Elo ratings ($\Delta^{(1)}$), and Figure~\ref{Fig2b} if it is based on the winning probability of the home team ($\Delta^{(2)}$). Although all matches are decided by a penalty shootout, the database is quite heterogeneous, the winning probability of the home team lies between 40\% and 60\% in less than 30\% of the sample.

The first covariate is being the first-mover; $f_h = 1$ ($f_h = -1$) if the home (away) team kicks the first penalty.
The second covariate is the comeback team; $c_h = 1$ ($c_h = -1$) if the home (away) team scores more goals or the last goal, while $c_h = 0$ if neither team scores more goals or the last goal. This choice ensures that a positive coefficient of these variables corresponds to a positive effect, whose extent is reflected by the absolute value of the coefficient.
There is no need to control for the match venue; the possible home (dis)advantage appears in the constant.

As a robustness check, two alternative definitions of the comeback team are considered. In the first, $c_h = 1$ ($c_h = -1$) if the home (away) team scores both more goals and the last goal. In the second, $c_h$ equals the goal difference from the perspective of the home team, similar to the method of \citet{Krumer2021b}.

To summarise, the following equation is estimated:
\[
P_h = \frac{1}{1 + e^{- \left( \beta_0 + \beta_1 \Delta + \beta_2 f_h + \beta_3 c_h \right)}},
\]
where $P_h$ is the probability that the home team wins the penalty shootout, and $\Delta$ is the strength difference of the home and the away team.
We consider 12 different regressions, depending on the definition of home venue (extended/restricted), the measure of strength difference (Elo rating/winning probability), and the momentum effect (ignored: $\beta_3 = 0$/more goals scored/last goal scored).

\cite{WunderlichBergeMemmertRein2020} warn that, although the logistic regression provides information about the connection between team strength and the outcome of the penalty shootout, this is an in-sample approach, which may suffer from overfitting. Since our results are strongly insignificant (see Section~\ref{Sec5}), we are not worried about this potential bias.

\section{Results} \label{Sec5}

Section~\ref{Sec51} investigates potential success factors in penalty shootouts by binomial tests, that is, whether teams in a given category win significantly more/less often than 50\% compared to teams in the other category. Section~\ref{Sec52} considers more dimensions simultaneously via logistic regressions. In both cases, sensitivity analyses are also conducted.

\subsection{Partial analysis} \label{Sec51}

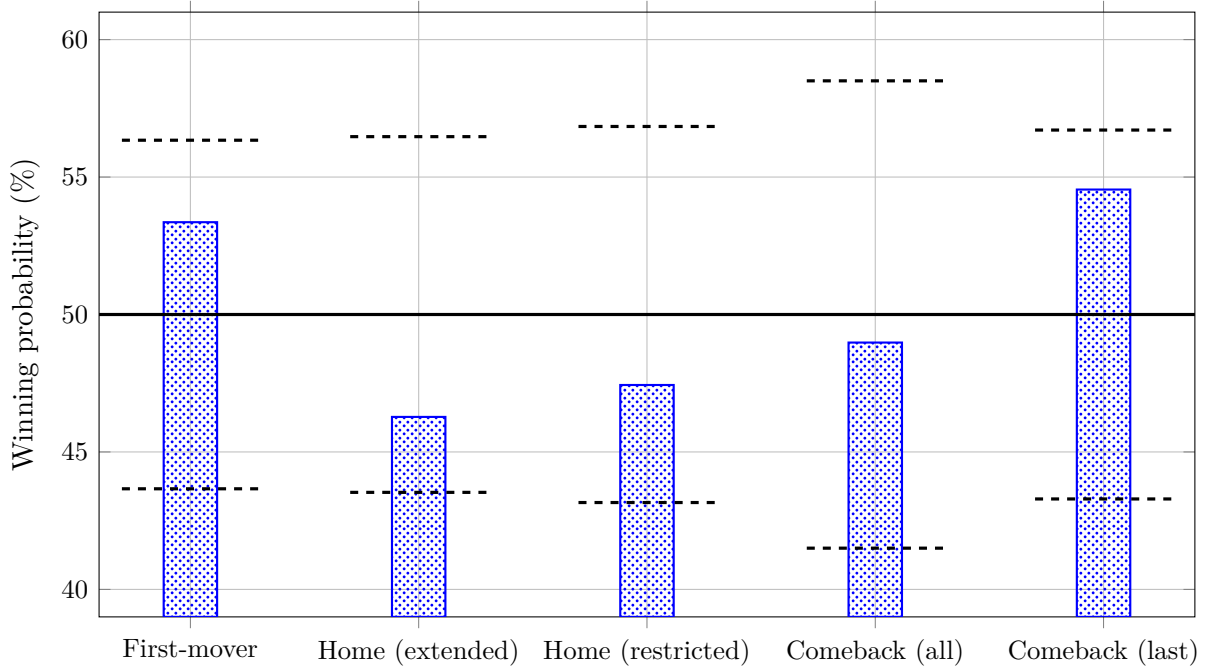
\begin{figure}[t!]
\centering

\begin{tikzpicture}
\begin{axis}[
width = \textwidth, 
height = 0.6\textwidth,
xmajorgrids,
ymajorgrids,
symbolic x coords = {First-mover,Home (extended),Home (restricted),Comeback (all),Comeback (last)},
xtick = data,
ylabel = {Winning probability (\%)},
ylabel style = {align=center, font=\small},
ymin = 39,
ymax = 61,
ybar,
bar width = 20pt,
yticklabel style = {/pgf/number format/fixed,/pgf/number format/precision=5},
]
\addplot [blue, thick, pattern = crosshatch dots, pattern color = blue] coordinates{
(First-mover,53.3582089552238)
(Home (extended),46.2745098039215)
(Home (restricted),47.4358974358974)
(Comeback (all),48.9795918367346)
(Comeback (last),54.5454545)
};

\coordinate (A) at (axis cs:{[normalized]-0.3},43.66);
\coordinate (B) at (axis cs:{[normalized]+0.3},43.66);
\coordinate (C) at (axis cs:{[normalized]-0.3},56.34);
\coordinate (D) at (axis cs:{[normalized]+0.3},56.34);

\draw [black,very thick,dashed] (A) -- (B);
\draw [black,very thick,dashed] (C) -- (D);

\coordinate (A) at (axis cs:{[normalized]0.7},43.53);
\coordinate (B) at (axis cs:{[normalized]1.3},43.53);
\coordinate (C) at (axis cs:{[normalized]0.7},56.47);
\coordinate (D) at (axis cs:{[normalized]1.3},56.47);

\draw [black,very thick,dashed] (A) -- (B);
\draw [black,very thick,dashed] (C) -- (D);

\coordinate (A) at (axis cs:{[normalized]1.7},43.16);
\coordinate (B) at (axis cs:{[normalized]2.3},43.16);
\coordinate (C) at (axis cs:{[normalized]1.7},56.84);
\coordinate (D) at (axis cs:{[normalized]2.3},56.84);

\draw [black,very thick,dashed] (A) -- (B);
\draw [black,very thick,dashed] (C) -- (D);

\coordinate (A) at (axis cs:{[normalized]2.7},41.5);
\coordinate (B) at (axis cs:{[normalized]3.3},41.5);
\coordinate (C) at (axis cs:{[normalized]2.7},58.5);
\coordinate (D) at (axis cs:{[normalized]3.3},58.5);

\draw [black,very thick,dashed] (A) -- (B);
\draw [black,very thick,dashed] (C) -- (D);

\coordinate (A) at (axis cs:{[normalized]3.7},43.29);
\coordinate (B) at (axis cs:{[normalized]4.3},43.29);
\coordinate (C) at (axis cs:{[normalized]3.7},56.71);
\coordinate (D) at (axis cs:{[normalized]4.3},56.71);

\draw [black,very thick,dashed] (A) -- (B);
\draw [black,very thick,dashed] (C) -- (D);


\draw [black,very thick] ({rel axis cs:0,0} |- {axis cs:First-mover,50}) -- ({axis cs:{Comeback (last)},50} -| {rel axis cs:1,0});

\end{axis}
\end{tikzpicture}

\captionsetup{justification=centerfirst}
\caption{Binomial tests: first-mover advantage, home advantage, momentum \\ \vspace{0.2cm}
\footnotesize{\emph{Note}: The dashed lines show the bounds of the 95\% confidence interval around no advantage of 50\% in a two-sided test.}}
\label{Fig3}
\end{figure}


Figure~\ref{Fig3} shows the results of five binomial tests that examine the effect of the shooting sequence, the match venue according to two definitions (mild and strict) of home field, as well as psychological momentum according to two assumptions again. None of them is significant even at the 10\% level. Thus, our study joins the set of papers that find no impact of the shooting order, and are fully in line with the literature on home advantage in penalty shootouts. In contrast to \citet[Table~4]{Krumer2021b}, comeback teams do not win more than 50\% of penalty shootouts.
These calculations can be reproduced on the basis of Table~\ref{Table_A1} in Appendix~\ref{Sec_A2}, which also reports the associated $p$-values.

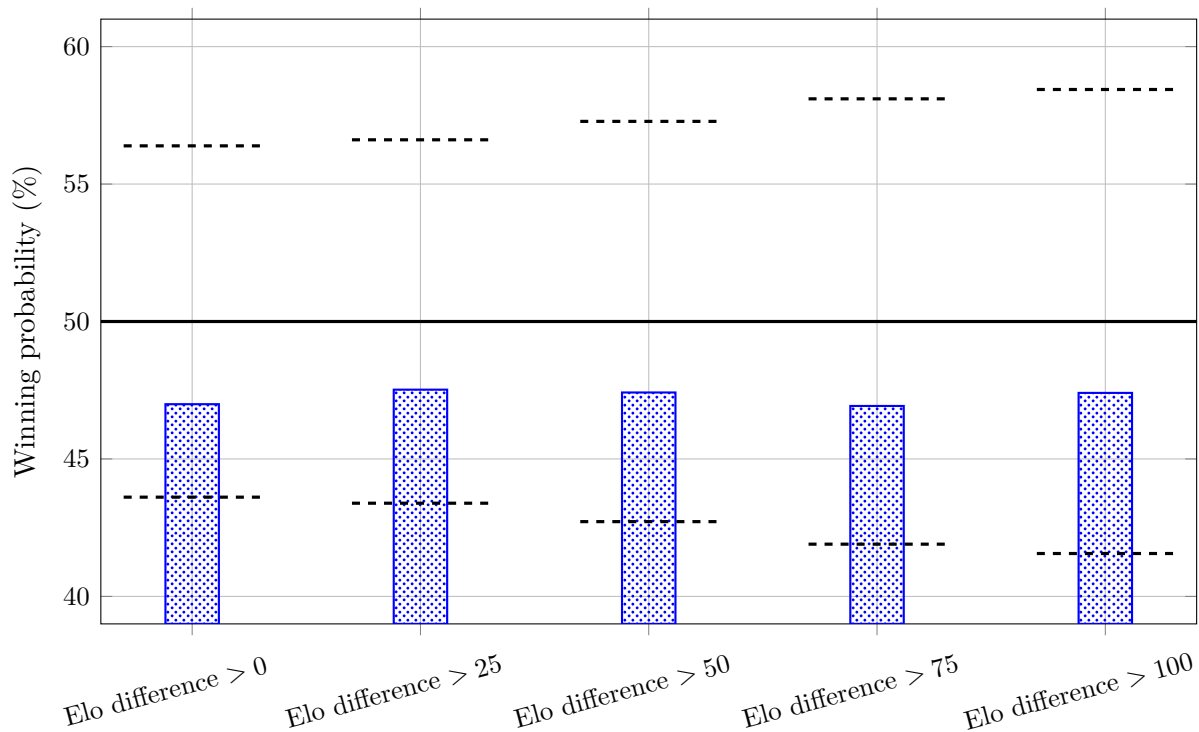
\begin{figure}[t!]
\centering

\begin{tikzpicture}
\begin{axis}[
width = \textwidth, 
height = 0.6\textwidth,
xmajorgrids,
ymajorgrids,
symbolic x coords = {Elo difference $> 0$,Elo difference $> 25$,Elo difference $> 50$,Elo difference $> 75$,Elo difference $> 100$},
xtick = data,
xticklabel style = {rotate=15, anchor=north, xshift=-15pt, yshift=-10pt},
ylabel = {Winning probability (\%)},
ylabel style = {align=center, font=\small},
ymin = 39,
ymax = 61,
ybar,
bar width = 20pt,
yticklabel style = {/pgf/number format/fixed,/pgf/number format/precision=5},
]
\addplot [blue, thick, pattern = crosshatch dots, pattern color = blue] coordinates{
(Elo difference $> 0$,46.9924812030075)
(Elo difference $> 25$,47.5206611570247)
(Elo difference $> 50$,47.4178403755868)
(Elo difference $> 75$,46.9273743016759)
(Elo difference $> 100$,47.4025974025974)
};

\coordinate (A) at (axis cs:{[normalized]-0.3},43.61);
\coordinate (B) at (axis cs:{[normalized]+0.3},43.61);
\coordinate (C) at (axis cs:{[normalized]-0.3},56.39);
\coordinate (D) at (axis cs:{[normalized]+0.3},56.39);

\draw [black,very thick,dashed] (A) -- (B);
\draw [black,very thick,dashed] (C) -- (D);

\coordinate (A) at (axis cs:{[normalized]0.7},43.39);
\coordinate (B) at (axis cs:{[normalized]1.3},43.39);
\coordinate (C) at (axis cs:{[normalized]0.7},56.61);
\coordinate (D) at (axis cs:{[normalized]1.3},56.61);

\draw [black,very thick,dashed] (A) -- (B);
\draw [black,very thick,dashed] (C) -- (D);

\coordinate (A) at (axis cs:{[normalized]1.7},42.72);
\coordinate (B) at (axis cs:{[normalized]2.3},42.72);
\coordinate (C) at (axis cs:{[normalized]1.7},57.28);
\coordinate (D) at (axis cs:{[normalized]2.3},57.28);

\draw [black,very thick,dashed] (A) -- (B);
\draw [black,very thick,dashed] (C) -- (D);

\coordinate (A) at (axis cs:{[normalized]2.7},41.9);
\coordinate (B) at (axis cs:{[normalized]3.3},41.9);
\coordinate (C) at (axis cs:{[normalized]2.7},58.1);
\coordinate (D) at (axis cs:{[normalized]3.3},58.1);

\draw [black,very thick,dashed] (A) -- (B);
\draw [black,very thick,dashed] (C) -- (D);

\coordinate (A) at (axis cs:{[normalized]3.7},41.56);
\coordinate (B) at (axis cs:{[normalized]4.3},41.56);
\coordinate (C) at (axis cs:{[normalized]3.7},58.44);
\coordinate (D) at (axis cs:{[normalized]4.3},58.44);

\draw [black,very thick,dashed] (A) -- (B);
\draw [black,very thick,dashed] (C) -- (D);

\draw [black,very thick] ({rel axis cs:0,0} |- {axis cs:Elo difference $> 0$,50}) -- ({axis cs:Elo difference $> 100$,50} -| {rel axis cs:1,0});

\end{axis}
\end{tikzpicture}

\captionsetup{justification=centerfirst}
\caption{Binomial tests: team strength \\ \vspace{0.2cm}
\footnotesize{\emph{Note}: The dashed lines show the bounds of the 95\% confidence interval around no advantage of 50\% in a two-sided test.}}
\label{Fig4}
\end{figure}


Figure~\ref{Fig4} focuses on the winning probability of the stronger team based on Football Club Elo Ratings. Five definitions of the favourite are used by gradually restricting the sample according to the minimal difference in the Elo ratings of the two teams. However, even teams that have at least 100 points more than their opponents fail to win 50\% of their penalty shootouts.
These calculations can be reproduced on the basis of Table~\ref{Table_A2} in Appendix~\ref{Sec_A2}, which also reports the associated $p$-values.
The lack of positive impact is in stark contrast to the findings of the previous literature: all existing papers \citep{ArrondelDuhautoisLaslier2019, Krumer2020b, WunderlichBergeMemmertRein2020, Pipke2025, vanHemertvanderKamp2026} demonstrate a significantly higher winning probability for a ``sufficiently'' stronger team.

As we have seen in Section~\ref{Sec3}, the number of penalty shootouts has substantially increased from the 2020/21 season. Thus, all binomial tests have also been conducted for the first 20 and the last six seasons, respectively. According to Tables~\ref{Table_A1}--\ref{Table_A2} in the Appendix, the conclusions remain unchanged if we focus on the beginning or the end of the dataset. The only significant value at the level of 5\% (but not at the level of 1\%) appears for home advantage based on the extended dataset: home teams are less likely to win a penalty shootout than their opponents between 2020 and 2025 if the matches played behind closed doors are taken into account (Table~\ref{Table_A1}). In these 20 one-legged matches, only seven were won by the home team. This can be explained by choking under pressure \citep{Harb-WuKrumer2019}, which was probably stronger than usual despite the absence of spectators, as, in contrast to the standard rule, the randomly drawn home team did not play away before. The marginally significant effect disappears when only proper two-legged matches are considered under the restricted definition of home venue (Table~\ref{Table_A1}).
Tables~\ref{Table_A3} and \ref{Table_A4} reinforce the insignificant impact of all factors considered by distinguishing the penalty shootouts played in the qualification and knockout phases.

Even though the binomial tests do not indicate a significant deviation from the theoretically expected 50\%, this might be caused by a small sample size. Furthermore, a failure to reject the null hypothesis rules out only sufficiently large effects. Therefore, it is worth noting that a first-mover advantage can be observed to some extent, especially between 2000 and 2019, when the team kicking the first penalty had a higher probability of winning by more than 10 percentage points. On the other hand, while home teams won more penalty shootouts in the first 19 seasons, they might be disadvantaged recently: since 2020, the winning probability of the home team is lower by about 30\% compared to the away team. Last but not least, a slight but consistent comeback team advantage is seen if the comeback team is identified by scoring the last goal, which is a novel definition in the literature. In particular, these teams enjoy an advantage of 9 (based on the whole sample) and 14 (in the last six seasons) percentage points over their opponents, albeit this is still statistically insignificant.
Interestingly, the team kicking the first penalty or scoring the last goal is more likely to win---by about five percentage points---in matches played in the knockout stage.

\subsection{Multi-dimensional analysis} \label{Sec52}

\begin{table}[t!]
  \centering
  \caption{Logistic regression models for predicting success in penalty shootouts}
  \label{Table2}

\begin{subtable}{\textwidth}
  \centering
  \caption{Extended definition of home-away matches}
  \label{Table2a}
\begin{threeparttable}
    \begin{tabularx}{1\textwidth}{lCCCCCC} \toprule \hiderowcolors
    Measure of strength & \multicolumn{3}{c}{Elo difference $\Delta^{(1)}$} & \multicolumn{3}{c}{Home win probability $\Delta^{(2)}$} \\ \midrule
    \multirow{2}[0]{*}{Constant ($\beta_0$)} & $-$0.147 & $-$0.142 & $-$0.172 & $-$0.146 & $-$0.129 & $-$0.229 \\
          & (0.127) & (0.135) & (0.129) & (0.312) & (0.329) & (0.319) \\
    \multirow{2}[0]{*}{Strength difference ($\beta_1$)} & 0.019 & 0.017 & 0.034 & $-$0.009 & $-$0.028 & 0.107 \\
          & (0.071) & (0.073) & (0.072) & (0.595) & (0.605) & (0.602) \\
    \multirow{2}[0]{*}{First-mover ($\beta_2$)} & 0.121 & 0.124 & 0.106 & 0.119 & 0.122 & 0.104 \\
          & (0.127) & (0.128) & (0.128) & (0.127) & (0.129) & (0.128) \\
    \multirow{2}[0]{*}{Comeback (all goals, $\beta_3$)} & \multirow{2}[0]{*}{---} & $-$0.021 & \multirow{2}[0]{*}{---} & \multirow{2}[0]{*}{---} & $-$0.031 & \multirow{2}[0]{*}{---} \\
          &       & (0.185) &       &       & (0.184) &  \\
    \multirow{2}[0]{*}{Comeback (last goal, $\beta_3$)} & \multirow{2}[0]{*}{---} & \multirow{2}[0]{*}{---} & 0.202 & \multirow{2}[0]{*}{---} & \multirow{2}[0]{*}{---} & 0.195 \\
          &       &       & (0.140) &       &       & (0.140) \\ \midrule
    Observations & 253   & 253   & 253   & 253   & 253   & 253 \\ \bottomrule
    \end{tabularx}
\begin{tablenotes} \footnotesize
\item
The dependent variable is 1 (0) if the home (away) team won the penalty shootout. Each column represents a separate regression.
\item
Standard errors are in parentheses. Significance: * $p < 5\%$; ** $p < 1\%$; *** $p < 0.1\%$.
\item
Elo difference $\Delta^{(1)}$ is the Elo of the home team minus the Elo of the away team divided by 100.
\end{tablenotes}
\end{threeparttable}
\end{subtable}

\vspace{0.5cm}
\begin{subtable}{\textwidth}
  \centering
  \caption{Restricted definition of home-away matches}
  \label{Table2b}
\begin{threeparttable}
    \begin{tabularx}{1\textwidth}{lCCCCCC} \toprule \hiderowcolors
    Measure of strength & \multicolumn{3}{c}{Elo difference $\Delta^{(1)}$} & \multicolumn{3}{c}{Home win probability $\Delta^{(2)}$} \\ \midrule
    \multirow{2}[0]{*}{Constant ($\beta_0$)} & $-$0.103 & $-$0.094 & $-$0.126 & $-$0.046 & $-$0.020 & $-$0.124 \\
          & (0.132) & (0.141) & (0.134) & (0.331) & (0.35) & (0.339) \\
    \multirow{2}[0]{*}{Strength difference ($\beta_1$)} & 0.005 & 0.002 & 0.020 & $-$0.120 & $-$0.149 & $-$0.008 \\
          & (0.075) & (0.077) & (0.076) & (0.628) & (0.641) & (0.638) \\
    \multirow{2}[0]{*}{First-mover ($\beta_2$)} & 0.037 & 0.041 & 0.027 & 0.033 & 0.038 & 0.023 \\
          & (0.132) & (0.134) & (0.133) & (0.133) & (0.134) & (0.133) \\
    \multirow{2}[0]{*}{Comeback (all goals, $\beta_3$)} & \multirow{2}[0]{*}{---} & $-$0.033 & \multirow{2}[0]{*}{---} & \multirow{2}[0]{*}{---} & $-$0.043 & \multirow{2}[0]{*}{---} \\
          &       & (0.186) &       &       & (0.185) &  \\
    \multirow{2}[0]{*}{Comeback (last goal, $\beta_3$)} & \multirow{2}[0]{*}{---} & \multirow{2}[0]{*}{---} & 0.164 & \multirow{2}[0]{*}{---} & \multirow{2}[0]{*}{---} & 0.157 \\
          &       &       & (0.145) &       &       & (0.145) \\ \midrule
    Observations & 232   & 232   & 232   & 232   & 232   & 232 \\ \bottomrule
    \end{tabularx}
\begin{tablenotes} \footnotesize
\item
The dependent variable is 1 (0) if the home (away) team won the penalty shootout. Each column represents a separate regression.
\item
Standard errors are in parentheses. Significance: * $p < 5\%$; ** $p < 1\%$; *** $p < 0.1\%$.
\item
Elo difference $\Delta^{(1)}$ is the Elo of the home team minus the Elo of the away team divided by 100.
\end{tablenotes}
\end{threeparttable}
\end{subtable}
\end{table}

Table~\ref{Table2} investigates the effect of strength difference between the teams on the probability of winning a penalty shootout by logistic regressions, similar to \citet{WunderlichBergeMemmertRein2020} and \citet{vanHemertvanderKamp2026}. We study only matches with a home and an away team (Table~\ref{Table2a} uses an extended and Table~\ref{Table2b} uses a restricted definition), and control for first-mover advantage.
Crucially, no significant coefficient can be found. Nonetheless, the constant is always negative in line with the results of binomial tests in Figure~\ref{Fig3}, indicating a potential home disadvantage. Albeit insignificant, the coefficient $\beta_2$ is always positive, while $\beta_3$ is negative/positive when the momentum effect is identified by scoring all goals/the last goal, as expected from the binomial tests. The coefficient $\beta_1$ of strength difference also remains robustly insignificant.

As mentioned in Section~\ref{Sec4}, we have carried out analogous logistic regression such that the comeback team is identified by both scoring more goals and scoring the last goal, as well as by controlling for goal difference (home team goals minus away team goals) instead of the trichotomous variable of scoring more goals, analogous to \citet{Krumer2021b}. These results are reported in Table~\ref{Table_A5}; no estimation contains any significant coefficient, even at the 10\% level.

\begin{figure}[t!]
\centering

\begin{subfigure}{\textwidth}
\centering
\caption{Strength difference $\Delta^{(1)}$: Elo rating of the home team minus Elo rating of the away team}
\label{Fig5a}

\begin{tikzpicture}
\begin{axis}[
width = \textwidth, 
height = 0.6\textwidth,
xmajorgrids,
xlabel = {Starting year of the first season in the estimation},
xlabel style = {align=center, font=\small},
x tick label style={/pgf/number format/.cd,set thousands separator={}},
ylabel = {$p$-value},
ylabel style = {align=center, font=\small},
xmin = 1999.5,
xmax = 2024.5,
ymin = 0.002,
ymax = 1.1,
ymode = log,
log ticks with fixed point,
y tick label style = {/pgf/number format/.cd,fixed,fixed zerofill,precision=2},
legend style = {font=\small,at={(-0.03,-0.15)},anchor=north west,legend columns=3},
legend entries = {$p$-value of the constant$\qquad$, $p$-value of $\beta_1$ for $\Delta^{(1)} \qquad$, $p$-value of $\beta_2$ for first-mover},
]
\addplot [red, mark = o, only marks] coordinates{
(2000,0.246034242016804)
(2001,0.22351205136626)
(2002,0.187429698550177)
(2003,0.133685706864305)
(2004,0.122590804580426)
(2005,0.0996480746109969)
(2006,0.071879663056067)
(2007,0.0690298402438497)
(2008,0.0798671295886919)
(2009,0.147760511037447)
(2010,0.109102178025157)
(2011,0.192899484632372)
(2012,0.169530954782397)
(2013,0.155126471851946)
(2014,0.211082845336361)
(2015,0.0816608695165328)
(2016,0.0728719082740687)
(2017,0.062202095841565)
(2018,0.050164818878209)
(2019,0.0458611531903265)
(2020,0.0320127319781453)
(2021,0.0629814077560799)
(2022,0.0241557847959861)
(2023,0.0757261675778356)
(2024,0.0143345457022534)
};
\addplot [blue, mark = diamond, only marks] coordinates{
(2000,0.789211853633348)
(2001,0.741585242121477)
(2002,0.862463753721107)
(2003,0.777948921516652)
(2004,0.599641606599857)
(2005,0.716517374229055)
(2006,0.810256106675068)
(2007,0.839368763668942)
(2008,0.970207820495698)
(2009,0.989782033868229)
(2010,0.989543734141545)
(2011,0.85863517654534)
(2012,0.707538298809854)
(2013,0.723750332493594)
(2014,0.770508311654271)
(2015,0.658885031367765)
(2016,0.916815757591383)
(2017,0.943564461760469)
(2018,0.712079361853695)
(2019,0.693611090716134)
(2020,0.829794028277013)
(2021,0.768894967373651)
(2022,0.244982024882177)
(2023,0.0895200092789476)
(2024,0.0149406612973695)
};
\addplot [ForestGreen, mark = star, only marks] coordinates{
(2000,0.338381993798945)
(2001,0.370811516405563)
(2002,0.272133201462526)
(2003,0.301255438011075)
(2004,0.328218052974974)
(2005,0.450610067505414)
(2006,0.537993208638626)
(2007,0.629732228196978)
(2008,0.846094827402683)
(2009,0.873673045977289)
(2010,0.864384829468115)
(2011,0.854611585063609)
(2012,0.784784296032207)
(2013,0.768424485794373)
(2014,0.702955925771038)
(2015,0.84756444204663)
(2016,0.739125665120957)
(2017,0.908399292598779)
(2018,0.990771080743086)
(2019,0.805389592234776)
(2020,0.940808629058864)
(2021,0.306597791530054)
(2022,0.162004001694096)
(2023,0.0406706618668239)
(2024,0.0033483129342469)
};

\draw [black,thick,dotted] (\pgfkeysvalueof{/pgfplots/xmin},0.05) -- (\pgfkeysvalueof{/pgfplots/xmax},0.05);
\draw [black,thick,dashed] (\pgfkeysvalueof{/pgfplots/xmin},0.01) -- (\pgfkeysvalueof{/pgfplots/xmax},0.01);
\end{axis}
\end{tikzpicture}
\end{subfigure}

\vspace{0.25cm}
\begin{subfigure}{\textwidth}
\centering
\caption{Strength difference $\Delta^{(2)}$: winning probability of the home team based on Elo ratings}
\label{Fig5b}

\begin{tikzpicture}
\begin{axis}[
width = \textwidth, 
height = 0.6\textwidth,
xmajorgrids,
xlabel = {Starting year of the first season in the estimation},
xlabel style = {align=center, font=\small},
x tick label style={/pgf/number format/.cd,set thousands separator={}},
ylabel = {$p$-value},
ylabel style = {align=center, font=\small},
xmin = 1999.5,
xmax = 2024.5,
ymin = 0.002,
ymax = 1.1,
ymode = log,
log ticks with fixed point,
y tick label style = {/pgf/number format/.cd,fixed,fixed zerofill,precision=2},
legend style = {font=\small,at={(-0.03,-0.15)},anchor=north west,legend columns=3},
legend entries = {$p$-value of the constant$\qquad$, $p$-value of $\beta_1$ for $\Delta^{(2)} \qquad$, $p$-value of $\beta_2$ for first-mover},
]
\addplot [red, mark = o, only marks] coordinates{
(2000,0.639211644007208)
(2001,0.579459805640552)
(2002,0.658092029056949)
(2003,0.525521080375706)
(2004,0.363920379211481)
(2005,0.425966930388832)
(2006,0.459792097279945)
(2007,0.480137030614006)
(2008,0.595726390731066)
(2009,0.747684267079714)
(2010,0.695490299967044)
(2011,0.64136563893871)
(2012,0.489267322383071)
(2013,0.47491032825609)
(2014,0.553926903209921)
(2015,0.347120096616617)
(2016,0.472834078067544)
(2017,0.464395678674535)
(2018,0.299136788547295)
(2019,0.273804130228331)
(2020,0.342595468525145)
(2021,0.714295227663497)
(2022,0.779205320596628)
(2023,0.278175867119267)
(2024,0.058481256945761)
};
\addplot [blue, mark = diamond, only marks] coordinates{
(2000,0.988487129921784)
(2001,0.957984137266154)
(2002,0.911421784662417)
(2003,0.991046989098463)
(2004,0.784655118647747)
(2005,0.901429164843461)
(2006,0.993072136158303)
(2007,0.96073490388041)
(2008,0.826673192948629)
(2009,0.75752587591603)
(2010,0.759094998596224)
(2011,0.916291057397355)
(2012,0.914698341176731)
(2013,0.908896046879056)
(2014,0.942174634751192)
(2015,0.813196540514799)
(2016,0.994134492761833)
(2017,0.996621731072269)
(2018,0.770527869590376)
(2019,0.726222047081005)
(2020,0.902597371092431)
(2021,0.735125440123907)
(2022,0.257303964501728)
(2023,0.0854469719996792)
(2024,0.0117943013275167)
};
\addplot [ForestGreen, mark = star, only marks] coordinates{
(2000,0.349921109378324)
(2001,0.381368883415952)
(2002,0.280971958798537)
(2003,0.308672235556905)
(2004,0.333436438426492)
(2005,0.457087080437383)
(2006,0.547322907017983)
(2007,0.640968508764409)
(2008,0.863150272312623)
(2009,0.897509161509691)
(2010,0.890143495743045)
(2011,0.878613915721338)
(2012,0.807200459570741)
(2013,0.790422391077021)
(2014,0.722736656643058)
(2015,0.862911022453807)
(2016,0.752820457604415)
(2017,0.916143309785102)
(2018,0.991604064476933)
(2019,0.8037107378797)
(2020,0.948003522453967)
(2021,0.300943025186392)
(2022,0.161209590820986)
(2023,0.0391247624649532)
(2024,0.00295692234562096)
};

\draw [black,thick,dotted] (\pgfkeysvalueof{/pgfplots/xmin},0.05) -- (\pgfkeysvalueof{/pgfplots/xmax},0.05);
\draw [black,thick,dashed] (\pgfkeysvalueof{/pgfplots/xmin},0.01) -- (\pgfkeysvalueof{/pgfplots/xmax},0.01);
\end{axis}
\end{tikzpicture}
\end{subfigure}

\captionsetup{justification=centerfirst}
\caption{Logistic regressions: Sensitivity of $p$-values to the first season of the sample \\ \vspace{0.2cm}
\footnotesize{\emph{Note}: The dotted (dashed) line shows significance at the 5\% (1\%) level.}}
\label{Fig5}
\end{figure}


Last but not least, we have assessed whether team strength could have a significant effect by focusing on the recent seasons. Hence, 25-25 logistic regressions are estimated by controlling for strength difference and first-mover effect with the extended definition of home field: the sample begins in the 20$x$/20$x$+1 season and ends in 2025. Figure~\ref{Fig5} plots the corresponding $p$-values for the coefficients $\beta_0$, $\beta_1$, and $\beta_2$.
The null hypothesis of equal winning probabilities can be rejected only when all seasons before 2023/24 are ignored---but then the sample size is not more than 64. Furthermore, $\beta_1$ is significant only if the last one or two season(s) are considered together with the first half of the 2025/26 season. Even then, both $\beta_1$ and $\beta_2$ are negative, which indicates rather a statistical anomaly caused by inadequate sample size.
Consequently, in contrast to domestic cups, stronger teams are not expected to win more penalty shootouts in UEFA club competitions. 

\section{Discussion and concluding remarks} \label{Sec6}

At first sight, our findings may seem to contradict results from the previous literature. The detailed comparison of the estimations by \citet{ApesteguiaPalacios-Huerta2010} and \citet{KocherLenzSutter2012} uncovers that the discrepancy is often caused by sampling issues, see \citet{KocherLenzSutter2012}. The majority of our sample consists of matches played in the last six seasons, which can explain at least some differences.

Regarding first-mover advantage, the results are fully in line with the conclusions of recent studies on large samples \citep{Pipke2025, VollmerSchochBrandes2024}.
Similar to us, no previous work has found evidence for home advantage.
Even though \citet{Krumer2021b} reports a momentum effect for away (but not for home) teams, there are two important differences between the two analyses. First, \citet{Krumer2021b} measures momentum by the goal difference between the teams in regular time of the second leg; thus, it is not a binary variable as in our case. Second, the sample of \citet{Krumer2021b} contains matches played between the 1970/71 and 2017/18 seasons, and there is powerful evidence for a decreasing home advantage in club football over time \citep{BakerMcHale2015}. Therefore, away teams were more disadvantaged in matches played decades ago, hence, it is logical to assume that they enjoyed a stronger momentum effect if they managed to score more goals against their opponents.

In contrast to all existing studies, stronger teams are found not to win more often, which can be deemed counter-intuitive. However, we do not see such a problem after a more detailed analysis.
First, \citet{ArrondelDuhautoisLaslier2019} and \citet{Krumer2020b} find that teams from a higher-ranked division have a higher probability of winning in a penalty shootout. Our sample is more homogeneous from this perspective: teams from the highest-ranked European leagues do not play in the UEFA Champions League qualification since the 2018/19 season, and at most one team participates in the UEFA Europa League and UEFA (Europa) Conference League qualification. Furthermore, the UEFA Champions League qualification was separated into two paths in the 2009/10 season, such that champions from lower-ranked national associations are not allowed to play against non-champions from higher-ranked national associations \citep{Csato2022b}. Even though the knockout stage does not contain analogous constraints, the previous group stage/league phase can effectively remove weak clubs, guaranteeing that teams playing against each other are quite close in abilities. 
Second, although \citet{WunderlichBergeMemmertRein2020} report a significant impact for betting odds in domestic cups, its magnitude is relatively small, even extremely weaker teams have a probability of around 40\% to win the penalty shootout. UEFA club competitions are more prestigious and lucrative than national clubs, thus strong teams probably field their best squad. Therefore, if the strength difference between the opposing teams is substantial, the match is less likely to be decided by a penalty shootout in UEFA club competitions than in national cups.

Any penalty shootout is clearly a situation with high stakes. In such an environment, choking under pressure can threaten the stronger team, which is expected to win. Indeed, professional biathlon athletes in the top quartile of the ability distribution miss significantly more shots when competing in their home country compared to competing abroad \citep{Harb-WuKrumer2019}. In addition, task-specific practice could contribute to the success of penalty kicks \citep{NavarroMiyamotovanderKampMoryaSavelsberghRanvaud2013}. This is especially important because, according to \citet{Jordet2009}, players from countries with many international club titles or many internationally decorated players (such as England or Spain), spend less time preparing for their shots and, consequently, are less successful in kicking penalties than players from countries with a lower public status.

Nevertheless, it is important not to over-interpret our finding that stronger teams do not win in a penalty shootout with a higher probability than their weaker opponents.
First, the whole analysis is \emph{conditional} on matches reaching a penalty shootout. Team strength increases the probability of winning earlier \citep{Csato2024c}, which implies that shootouts involve more closely matched teams, as verified by recent empirical evidence in UEFA club competitions \citep{ForrestKosciolekTena2025}.
Second, we have focused on one measure of team strength, Football Club Elo Ratings. Using alternative measures of strength, such as UEFA club coefficients or market values, may change the results.

To conclude, the current paper has aimed to predict success in penalty shootouts in UEFA club competitions during the last 25 years. We have considered several possible factors, such as the shooting order, the field of the match, and psychological momentum, but none of them affects the probability of winning significantly. First in the literature, team strength is measured by Elo ratings instead of the division of the team or betting odds. This has led to a somewhat surprising finding: even though Elo ratings can forecast results in the UEFA Champions League relatively well, they seem to lose their predictive power if qualification is decided by penalty shootouts.


\section*{Acknowledgements}
\addcontentsline{toc}{section}{Acknowledgements}
\noindent
We appreciate the help of \emph{Mil\'an Kolonics} in data collection. \\
We are grateful to \emph{Alex Krumer} and \emph{Fabian Wunderlich} for their useful comments. \\
Two anonymous reviewers provided valuable remarks on earlier drafts. \\
The research was supported by the National Research, Development and Innovation Office under Grants Advanced 152220, FK 145838, and PD 153835, and by the J\'anos Bolyai Research Scholarship of the Hungarian Academy of Sciences.

\bibliographystyle{apalike}
\bibliography{All_references}

\clearpage
\setcounter{section}{0}
\renewcommand{\thesubsection}{A.\arabic{subsection}}

\setcounter{table}{0}
\renewcommand{\thetable}{A.\arabic{table}}

\section*{Appendix}
\addcontentsline{toc}{section}{Appendix}

\subsection{A mistake in Football Club Elo Ratings data} \label{Sec_A1}

Two Armenian football clubs, Ararat Yerevan (\url{http://clubelo.com/Ararat}) and Ararat-Armenia (\url{http://clubelo.com/Ararat-Armenia}), have similar names. Both teams have played in UEFA club competitions recently.

In the 2021/22 Armenian Premier League, Ararat-Armenia was the runner-up and qualified for the second qualifying round of the 2022/23 UEFA Europa Conference League. Ararat Yerevan was ranked fourth and qualified for the first qualifying round of the 2022/23 UEFA Europa Conference League. Ararat Yerevan lost in the first round against the North Macedonian club Shk\"endiye, while Ararat-Armenia lost in the second round---by penalty shootouts---against the Estonian club Paide Linnameeskond.
However, Football Club Elo Ratings registered both matches for Ararat Yerevan, see \url{http://clubelo.com/Ararat/Games/Latest}.
Thus, the Elo rating of Ararat-Armenia remains unknown in these matches.

In the 2022/23 Armenian Premier League, Ararat-Armenia was ranked third and qualified for the first qualifying round of the 2023/24 UEFA Europa Conference League. It won in the first round---by penalty shootouts---against the Albanian club Egnatia, and lost in the second round against the Greek club Aris.
However, Football Club Elo Ratings registered both matches for Ararat Yerevan, see \url{http://clubelo.com/Ararat/Games/Latest}.
Thus, the Elo rating of Ararat-Armenia remains unknown in these matches.

In the 2023/24 Armenian Premier League, Ararat-Armenia was ranked third and qualified for the second qualifying round of the 2024/25 UEFA Conference League. It won in the second round against the Moldavian club Zimbru Chi{\c s}in{\u a}u, and lost in the third round against the Hungarian club Pusk\'as Akad\'emia.
However, Football Club Elo Ratings registered both matches for Ararat Yerevan, see \url{http://clubelo.com/Ararat/Games/Latest}.
Thus, the Elo rating of Ararat-Armenia remains unknown in these matches.

In the 2024/25 Armenian Premier League, Ararat-Armenia was ranked second and qualified for the second qualifying round of the 2025/26 UEFA Conference League. It won in the second round against the Romanian club Universitatea Cluj, and lost in the third round against the Czech club Sparta Praha.
However, Football Club Elo Ratings registered both matches for Ararat Yerevan, see \url{http://clubelo.com/Ararat/Games/Latest}. 
Thus, the Elo rating of Ararat-Armenia remains unknown in these matches.

Therefore, we could not know the Elo rating of one team, Ararat-Armenia, in two penalty shootouts, which were ignored in the corresponding analyses.

More worryingly, the unreliability of the Elo ratings for Ararat Yerevan spreads to their (supposed) opponents, opponents of opponents, and so on, raising questions about the validity of Football Club Elo Ratings data. Unfortunately, we are not able to correct this error of the underlying database. We have informed the creator of Football Club Elo Ratings, \emph{Lars Schiefler}, about the problem but received no answer.

\clearpage
\subsection{Supplementary tables} \label{Sec_A2}

\begin{table}[ht!]
  \centering
  \caption{Results of binomial tests: first-mover advantage, home advantage, momentum}
  \label{Table_A1}
\centerline{
\begin{threeparttable}
    \rowcolors{1}{gray!20}{}
    \begin{tabularx}{1.15\textwidth}{l Ccc Ccc Ccc} \toprule \hiderowcolors
    Period & \multicolumn{3}{c}{2000/01--2025} & \multicolumn{3}{c}{2000/01--2019/20} & \multicolumn{3}{c}{2020/21--2025} \\
    Potential success factor & Total & Ratio & $p$-value & Total & Ratio & $p$-value & Total & Ratio & $p$-value \\ \bottomrule \showrowcolors
    First-mover & 268   & 53.4\% & 0.299 & 129   & 55.8\% & 0.218 & 139   & 51.1\% & 0.865 \\
    Home (extended data) & 255   & 46.3\% & 0.260 & 122   & 52.5\% & 0.651 & 133   & 40.6\% & 0.037* \\
    Home (restricted data) & 234   & 47.4\% & 0.472 & 121   & 52.9\% & 0.586 & 113   & 41.6\% & 0.090 \\
    Comeback (all goals) & 147   & 49.0\% & 0.869 & 75    & 53.3\% & 0.644 & 72    & 44.4\% & 0.410 \\
    Comeback (last goal) & 231   & 54.5\% & 0.188 & 108   & 51.9\% & 0.773 & 123   & 56.9\% & 0.149 \\ \toprule
    \end{tabularx}
\begin{tablenotes} \footnotesize
\item
Two-sided binomial tests. Significance: * $p < 5\%$; ** $p < 1\%$; *** $p < 0.1\%$.
\item
The column Total (Ratio) shows the number (proportion) of penalty shootouts won by the team indicated in the row.
\end{tablenotes}
\end{threeparttable}
}
\end{table}

\begin{table}[ht!]
  \centering
  \caption{Results of binomial tests: team strength}
  \label{Table_A2}
\centerline{
\begin{threeparttable}
    \rowcolors{1}{gray!20}{}
    \begin{tabularx}{1.15\textwidth}{l CCc CCc CCc} \toprule \hiderowcolors
    Period & \multicolumn{3}{c}{2000/01--2025} & \multicolumn{3}{c}{2000/01--2019/20} & \multicolumn{3}{c}{2020/21--2025} \\
    Scenario & Total & Ratio & $p$-value & Total & Ratio & $p$-value & Total & Ratio & $p$-value \\ \bottomrule \showrowcolors
    Elo difference $> 0$ & 266   & 47.0\% & 0.358 & 129   & 46.5\% & 0.481 & 137   & 47.4\% & 0.608 \\
    Elo difference $> 25$ & 242   & 47.5\% & 0.480 & 120   & 46.7\% & 0.523 & 122   & 48.4\% & 0.786 \\
    Elo difference $> 50$ & 213   & 47.4\% & 0.493 & 108   & 46.3\% & 0.501 & 105   & 48.6\% & 0.845 \\
    Elo difference $> 75$ & 179   & 46.9\% & 0.455 & 92    & 45.7\% & 0.466 & 87    & 48.3\% & 0.830 \\
    Elo difference $> 100$ & 154   & 47.4\% & 0.573 & 76    & 47.4\% & 0.731 & 78    & 47.4\% & 0.734 \\ \toprule
    \end{tabularx}
\begin{tablenotes} \footnotesize
\item
Two-sided binomial tests. Significance: * $p < 5\%$; ** $p < 1\%$; *** $p < 0.1\%$.
\item
The column Total (Ratio) shows the number (proportion) of penalty shootouts won by the favourite team.
\end{tablenotes}
\end{threeparttable}
}
\end{table}

\begin{table}[t!]
  \centering
  \caption{Results of binomial tests in different tournament stages: \\ first-mover advantage, home advantage, momentum}
  \label{Table_A3}
\centerline{
\begin{threeparttable}
    \rowcolors{1}{gray!20}{}
    \begin{tabularx}{1\textwidth}{l CCC CCC} \toprule \hiderowcolors
    Tournament phase & \multicolumn{3}{c}{Qualification} & \multicolumn{3}{c}{Knockout (including the final)} \\
        Potential success factor & Total & Ratio & $p$-value & Total & Ratio & $p$-value \\ \bottomrule \showrowcolors
    First-mover & 207   & 52.2\% & 0.578 & 61    & 57.4\% & 0.306 \\
    Home (extended data) & 204   & 46.1\% & 0.294 & 51    & 47.1\% & 0.780 \\
    Home (restricted data) & 184   & 47.3\% & 0.507 & 50    & 48.0\% & 0.888 \\
    Comeback (all goals) & 108   & 49.1\% & 0.923 & 39    & 48.7\% & 1.000 \\
    Comeback (last goal) & 176   & 53.4\% & 0.407 & 55    & 58.2\% & 0.281 \\ \toprule
    \end{tabularx}
\begin{tablenotes} \footnotesize
\item
Two-sided binomial tests. Significance: * $p < 5\%$; ** $p < 1\%$; *** $p < 0.1\%$.
\item
The column Total (Ratio) shows the number (proportion) of penalty shootouts won by the team indicated in the row.
\end{tablenotes}
\end{threeparttable}
}
\end{table}

\begin{table}[ht!]
  \centering
  \caption{Results of binomial tests in different tournament stages: team strength}
  \label{Table_A4}
\centerline{
\begin{threeparttable}
    \rowcolors{1}{gray!20}{}
    \begin{tabularx}{1\textwidth}{l CCC CCC} \toprule \hiderowcolors
    Tournament phase & \multicolumn{3}{c}{Qualification} & \multicolumn{3}{c}{Knockout (including the final)} \\
    Scenario & Total & Ratio & $p$-value & Total & Ratio & $p$-value \\ \bottomrule \showrowcolors
    Elo difference $> 0$ & 205   & 48.3\% & 0.675 & 61    & 42.6\% & 0.306 \\
    Elo difference $> 25$ & 188   & 49.5\% & 0.942 & 54    & 40.7\% & 0.220 \\
    Elo difference $> 50$ & 165   & 49.1\% & 0.876 & 48    & 41.7\% & 0.312 \\
    Elo difference $> 75$ & 145   & 46.9\% & 0.507 & 34    & 47.1\% & 0.864 \\
    Elo difference $> 100$ & 134   & 47.0\% & 0.546 & 20    & 50.0\% & 1.000 \\ \toprule
    \end{tabularx}
\begin{tablenotes} \footnotesize
\item
Two-sided binomial tests. Significance: * $p < 5\%$; ** $p < 1\%$; *** $p < 0.1\%$.
\item
The column Total (Ratio) shows the number (proportion) of penalty shootouts won by the favourite team.
\end{tablenotes}
\end{threeparttable}
}
\end{table}

\begin{table}[t!]
  \centering
  \caption{Logistic regression models for predicting success in penalty shootouts \\
  Robustness check with alternative definitions of the comeback team}
  \label{Table_A5}

\begin{subtable}{\textwidth}
  \centering
  \caption{Extended definition of home-away matches}
  \label{Table_A5a}
\begin{threeparttable}
    \begin{tabularx}{1\textwidth}{lCCCC} \toprule \hiderowcolors  
    Measure of strength & \multicolumn{2}{c}{Elo difference $\Delta^{(1)}$} & \multicolumn{2}{c}{Home win probability $\Delta^{(2)}$} \\ \midrule
    \multirow{2}[0]{*}{Constant ($\beta_0$)} & $-$0.158 & $-$0.160 & $-$0.168 & $-$0.168 \\
          & (0.133) & (0.135) & (0.327) & (0.325) \\
    \multirow{2}[0]{*}{Strength difference ($\beta_1$)} & 0.023 & 0.022 & 0.017 & 0.013 \\
          & (0.073) & (0.072) & (0.606) & (0.601) \\
    \multirow{2}[0]{*}{First-mover ($\beta_2$)} & 0.116 & 0.114 & 0.114 & 0.112 \\
          & (0.128) & (0.129) & (0.128) & (0.129) \\
    \multirow{2}[0]{*}{Comeback (double, $\beta_3$)} & 0.055 & \multirow{2}[0]{*}{---} & 0.044 & \multirow{2}[0]{*}{---} \\
          & (0.199) &       & (0.199) &  \\
    \multirow{2}[0]{*}{Comeback (goal diff., $\beta_3$)} & \multirow{2}[0]{*}{---} & 0.040 & \multirow{2}[0]{*}{---} & 0.034 \\
          &       & (0.140) &       & (0.139) \\ \midrule
    Observations & 253   & 253   & 253   & 253 \\ \bottomrule
    \end{tabularx}
\begin{tablenotes} \footnotesize
\item
The dependent variable is 1 (0) if the home (away) team won the penalty shootout. Each column represents a separate regression.
\item
Standard errors are in parentheses. Significance: * $p < 5\%$; ** $p < 1\%$; *** $p < 0.1\%$.
\item
Elo difference $\Delta^{(1)}$ is the Elo of the home team minus the Elo of the away team divided by 100.
\item
Comeback (double) is 1 ($-1$) if the home (away) team scored both more goals and the last goal.
\item
Comeback (goal diff.) is the number of goals scored minus the number of goals conceded for the home team.
\end{tablenotes}
\end{threeparttable}
\end{subtable}

\vspace{0.5cm}
\begin{subtable}{\textwidth}
  \centering
  \caption{Restricted definition of home-away matches}
  \label{Table_A5b}
\begin{threeparttable}
    \begin{tabularx}{1\textwidth}{lCCCC} \toprule \hiderowcolors  
    Measure of strength & \multicolumn{2}{c}{Elo difference $\Delta^{(1)}$} & \multicolumn{2}{c}{Home win probability $\Delta^{(2)}$} \\ \midrule
    \multirow{2}[0]{*}{Constant ($\beta_0$)} & $-$0.113 & $-$0.116 & $-$0.066 & $-$0.068 \\
          & (0.139) & (0.141) & (0.348) & (0.345) \\
    \multirow{2}[0]{*}{Strength difference ($\beta_1$)} & 0.009 & 0.009 & $-$0.096 & $-$0.098 \\
          & (0.077) & (0.076) & (0.642) & (0.636) \\
    \multirow{2}[0]{*}{First-mover ($\beta_2$)} & 0.033 & 0.030 & 0.030 & 0.027 \\
          & (0.134) & (0.135) & (0.134) & (0.135) \\
    \multirow{2}[0]{*}{Comeback (double, $\beta_3$)} & 0.047 & \multirow{2}[0]{*}{---} & 0.036 & \multirow{2}[0]{*}{---} \\
          & (0.200) &       & (0.200) &  \\
    \multirow{2}[0]{*}{Comeback (goal diff., $\beta_3$)} & \multirow{2}[0]{*}{---} & 0.037 & \multirow{2}[0]{*}{---} & 0.031 \\
          &       & (0.140) &       & (0.140) \\ \midrule
    Observations & 232   & 232   & 232   & 232 \\ \bottomrule
    \end{tabularx}
\begin{tablenotes} \footnotesize
\item
The dependent variable is 1 (0) if the home (away) team won the penalty shootout. Each column represents a separate regression.
\item
Standard errors are in parentheses. Significance: * $p < 5\%$; ** $p < 1\%$; *** $p < 0.1\%$.
\item
Elo difference $\Delta^{(1)}$ is the Elo of the home team minus the Elo of the away team divided by 100.
\item
Comeback (double) is 1 ($-1$) if the home (away) team scored both more goals and the last goal.
\item
Comeback (goal diff.) is the number of goals scored minus the number of goals conceded for the home team.
\end{tablenotes}
\end{threeparttable}
\end{subtable}
\end{table}

\end{document}